\documentclass[12pt,preprint]{aastex}

\newcommand{\rd}{{\rm d}}
\newcommand{\bF}{{\bf F}}
\newcommand{\bV}{{\bf V}}
\newcommand{\bJ}{{\bf J}}
\newcommand{\bW}{{\bf \Omega}}
\newcommand{\br}{{\bf r}}

\newcommand{\he}{{\hat{\bf e}}}
\newcommand{\hx}{{\hat{\bf x}}}
\newcommand{\hy}{{\hat{\bf y}}}
\newcommand{\hz}{{\hat{\bf z}}}
\newcommand{\bT}{{\bf T}}

\newcommand{\ba}{\begin{eqnarray}}
\newcommand{\ea}{\end{eqnarray}}

\newcommand{\lp}{\left(}
\newcommand{\rp}{\right)}
\newcommand{\lb}{\left[}
\newcommand{\rb}{\right]}
\newcommand{\ls}{\left<}
\newcommand{\rs}{\right>}

\begin{document}
\title{Spin-Kick Correlation in Neutron Stars: Alignment Conditions and Implications}
\author{Chen Wang\altaffilmark{1,2}, 
  Dong Lai\altaffilmark{2,1}, 
  J. L. Han\altaffilmark{1}}
\altaffiltext{1}{National Astronomical Observatories, Chinese Academy of
  Sciences, Jia 20 Datun Road, Chaoyang District, Beijing, 100012, China;
  wangchen@bao.ac.cn, hjl@bao.ac.cn}
\altaffiltext{2}{Department of Astronomy, Cornell University, 
Ithaca, NY 14853; dong@astro.cornell.edu}

\begin{abstract}
Recent observations of pulsar wind nebulae and radio polarization
profiles revealed a tendency of the alignment between the spin and
velocity directions in neutron stars. We study the condition for
spin-kick alignment using a toy model, in which the kick consists of
many off-centered, randomly-oriented thrusts.  Both analytical
considerations and numerical simulations indicate that spin-kick
alignment cannot be easily achieved if the proto-neutron star does not
possess some initial angular momentum, contrary to some previous
claims.  To obtain the observed spin-kick misalignment angle
distribution, the initial spin period of the neutron star must be
smaller than the kick timescale. Typically, an initial period of a
hundred milliseconds or less is required.

\end{abstract}

\keywords{neutron star --- pulsar kick}


\section{Introduction}

It is well known that pulsars have much larger space velocities than
their progenitors, implying a kick at neutron star (NS) birth (e.g.,
Lorimer et al.~1997; Arzoumanian et al.~2002; Chatterjee et al.~2005;
Hobbs et al.~2005; Winkler \& Petre~2006). The physical mechanism for
the kick, however, remains unclear (e.g., Lai 2004; Janka et
al.~2005).  One of the reasons that it has been difficult to constrain
various kick mechanisms is the lack of correlations between kick
velocity and the other properties of NSs. This situation has been
changed due to the recent high-resolution Chandra X-ray observations
of pulsar wind nebulae (e.g., Pavlov et al. 2000; Helfand et al. 2001;
Ng \& Romani 2004), which provided the evidence for spin-kick
alignment for several young pulsars (e.g, Lai et al.~2001;
Romani~2004; Wang et al.~2006)\footnote{A recent re-analysis of the
proper motion of the Crab pulsar (Ng \& Romani~2006) indicates that
the spin-kick misalignment angle is $26\arcdeg\pm3\arcdeg$}.

Another well-known method to determine the spin axis of pulsars is by
the linear polarization profile of radio emission.  If the
polarization profile could be described by rotating vector model
(RVM), one can constrain the projected spin axis by the polarization
angle at the center of the pulse. Previous attempts using this method
have yielded ambiguous results (e.g., Morris et al.~1979; Anderson \&
Lyne 1983; Deshpande et al.~1999) mainly because for many pulsars the
polarization profiles are not well described by the RVM (Weisberg et
al.~1999). With Parkes surveys (e.g., Manchester et al.~1996, 2001),
many well-calibrated polarization profiles became available, and more
pulsar rotation measures to eliminate the Faraday rotation effect
(e.g. Han et al.~2006). Moreover, proper motions for more than 200
pulsars have been determined (Hobbs et al.~2005).  By selecting
pulsars with well calibrated polarization and proper motion
measurements, Wang et al.~(2006) have obtained spin-kick misalignment
angle for 24 pulsars, and the data revealed a strong tendency of
spin-kick alignment. Johnston et al.~(2005) independently obtained
similar results for 25 pulsars based on different sample.

On the other hand, one can constrain NS kicks using the orbital
properties of NS binary systems (e.g., Dewey \& Cordes~1987; Fryer \&
Kalogera~1997; Willem et al.~2004; Thorsett et al.~2005).  In Wang et
al.~(2006), we obtained constraints on the kick magnitudes and
directions for various NS binaries, including double NS systems,
binaries with massive main-sequence companions, and binaries with
massive white-dwarf companions. We found that the kick velocity is
misaligned with the NS spin axis in a number of systems, and the NS
spin period (when available) in these systems is generally longer than
several hundreds milliseconds.

What is the implications of the apparent spin-kick alignment for many
pulsars?  One possibility which is widely discussed (e.g., Johnston et
al.~2005) comes from Spruit \& Phinney (1998). They suggested that the
initial spin of NS may originate from off-centered kicks even when the
proto-NS has no angular momentum. They further suggested that if one
imagine that the kick is composed of many random thrusts, then with
multiple thrusts, alignment may be easily achieved.

In this paper, we systematically study the condition of spin-kick
alignment using a toy model similar to that of Spruit \& Phinney
(1998).  We consider both the cases of zero and finite initial
proto-NS spin.  In \S 2 we introduce our toy model for kicks and
describe our simulation procedure. In \S3 we derive approximate but
analytic conditions for spin-kick alignment.  In \S 4, we present our
simulation results (especially the distribution of spin-kick
misalignment angle) under different initial conditions.  We find that,
consistently with our analytical estimate, without initial spin of the
proto-NS, it is difficult to achieve spin-kick alignment.  This is
contrary to some previous claims (e.g., Spruit \& Phinney~1998;
Johnston et al. 2005).  However, with sufficiently short initial spin
period (less than the timescale for each kick thrust), spin-kick
alignment can be achieved. We discuss the implications of our results
in \S5.



\section{A Toy Model for Kicks}

The basic equations governing the evolution of the center-of-mass
velocity $\bV$ and angular velocity $\bW$ of a proto-NS (mass $M$,
radius $R_{NS}$) are
\begin{eqnarray}
M\frac{\rd \bV}{{\rd} t} &=& \bF, \label{eq:Newton2}\\
\frac{\rd \bJ}{{\rd} t}  &=& \br \times \bF. \label{eq:am}
\end{eqnarray}
Here $\bF$ is the kick force, and $\br$ specifies the location where
the force is applied, $\bJ = I \bW$ is the angular momentum, with
$I=kMR_{\rm NS}^2$ the moment of inertia.  We adopt $M=1.4 M_\sun$,
$R_{\rm NS}=10\,{\rm km}$ and $k=0.4$.

We model the kick force on the NS as consisting of $n$ thrusts,
$\bF_i$, $i=1, 2, ...., n$, each has a duration $\tau_i$.  During each
thrust, we construct a ``temporary'' body frame (xyz) corotating with
the star so that $\bF_i$ is constant in this frame (see Fig.~1). We
specify the kick force $\bF_i$ by the magnitude $F_i$ ($=F$, the same
for all thrusts) and two angles $\alpha_i, \beta_i$.  We choose
($\alpha_i$, $\beta_i$) randomly distributed in the range of of
$0\arcdeg<\alpha_i<30\arcdeg$ and $0\arcdeg<\beta_i<360\arcdeg$.  Each
thrust acts at the position $\br_i$, which is specified by the
spherical coordinates $(r_i, \theta_i, \phi_i)$ in the nonrotating
frame (XYZ) at the beginning of the thrust. Note that choosing a
different range of $\alpha_i$ is equivalent to choosing different
$r_i$, as long as $r_i\sin\alpha_i$ remains the same. During the course
of each thrust (duration $\tau_i$), the $\br_i$ changes in the XYZ
frame as the body rotates.  In our simulation, we assume $\tau_i=\tau$
is the same for all thrusts.  Thus the total kick time is $T_{\rm
kick} = \sum_i\tau_i = n\tau$. We set the total momentum ${\cal P}
\equiv FT_{\rm kick}$ at the fixed range ${\cal P}=M{\cal
V}=M(500-2000)\,{\rm km\,s^{-1}}$ in all simulations. We fix
$r_i=30\,{\rm km}$ for all thrusts. For $(\theta_i,
\phi_i)$, we consider two possibilities: (i) The kick position is
randomly distributed on a sphere in the XYZ frame; (ii) The kick
position is randomly distributed on a sphere in the body frame of the
NS. Our simulation results reported in \S 4 refer to the first
case. We have found that the results of the second case are similar.

\begin{figure}
\begin{center}
\includegraphics[angle=-90, scale=0.6]{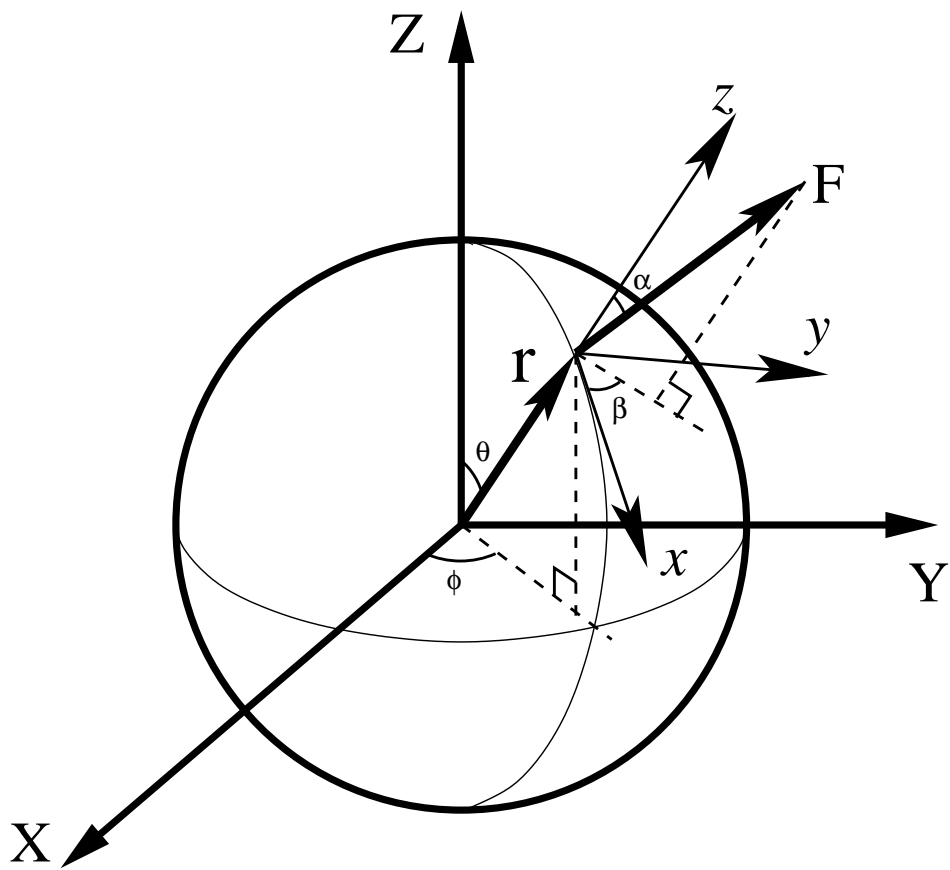}
\caption
     {Geometric model used in our simulations. XYZ is a nonrotating
      ``fixed'' frame centered at the neutron star. A thrust force
      $\bF_i$ is applied at the position $\br_i$ (specified by the
      spherical coordinates $r_i$, $\theta_i$, $\phi_i$. The body
      frame xyz is constructed with the $z$-axis along $\br_i$, and
      the $x$-axis in the meridional direction.  The direction of
      $\bF_i$ is specified by the two polar angles
      $\alpha_i,\beta_i$.}
\label{fig1}
\end{center}
\end{figure}

For each thrust, $\br_i$ and $\bF_i$ are fixed in the the
``temporary'' body frame, i.e., $\br_i=r_i\hz$, and
$\bF_i=F_i(\sin\alpha_i\cos\beta_i\hx+\sin\alpha_i\sin\beta_i\hy
+\cos\alpha_i\hz)$. To find the components of the force $\bF_i$ and
torque $\br_i\times\bF_i$ in the inertial frame (XYZ), we need to
solve for the time evolution of the body axes $\he$ ($=\hx$, $\hy$ or
$\hz$).  The (XYZ) frame and (xyz) frame are related by the rotation
matrix $\bT$:
\begin{eqnarray}
  \left( \begin{array}{c} X \\ Y \\ Z \end{array} \right) = 
  \bT \left( \begin{array}{c} x \\ y \\z \end{array} \right)
      = \left( \begin{array}{ccc} 
        T_{11} & T_{12} & T_{13} \\
        T_{21} & T_{22} & T_{23} \\
        T_{31} & T_{32} & T_{33}          
        \end{array} \right)
        \left( \begin{array}{c} x \\ y \\z \end{array} \right).
\end{eqnarray}
So we have the expressions of the body axes in the (XYZ) frame:
\begin{eqnarray}
\hx = \left( \begin{array}{c} T_{11} \\ T_{21} \\ T_{31} \end{array} \right), \quad
\hy = \left( \begin{array}{c} T_{12} \\ T_{22} \\ T_{32} \end{array} \right), \quad
\hz = \left( \begin{array}{c} T_{13} \\ T_{23} \\ T_{33} \end{array} \right). \label{eq:eexp}
\end{eqnarray}
At the beginning of each thrust
\begin{eqnarray}
  \bT = \left( \begin{array}{ccc} 
        \cos\theta_i\cos\phi_i & -\sin\phi_i & \sin\theta_i\cos\phi_i \\
        \cos\theta_i\sin\phi_i & \cos\phi_i& \sin\theta_i \sin\phi_i \\
        -\sin\theta_i      & 0         & \cos\theta_i         
        \end{array} \right).
\end{eqnarray}
The body axis $\he$ evolves according to 
\begin{eqnarray}
\frac{\rd \he}{\rd t}=\bW\times\he. \label{eq:eevo}
\end{eqnarray}
Substituting Eq.~(\ref{eq:eexp}) in Eq.~(\ref{eq:eevo}), we obtain the
evolution of each component of $\bT$, e.g., $\rd T_{11}/\rd t=\Omega_2
T_{31}-\Omega_3 T_{21}$, et al., where $\Omega_{1,2,3}$ are the three
components of $\bW$ in the XYZ frame.

Using Eqs.~(\ref{eq:Newton2})~--~(\ref{eq:eevo}), we can directly
simulate the movement and the rotation of the NS. Consider a star with
initial velocity $\bV_0=\bV_{\rm init}$ and angular velocity
$\bW_0=\bW_{\rm init}$.  Suppose it receives a thrust $\bF_1$ at a
random position $\br_1$ with duration $\tau_1$.  We use the 4th order
Runge-Kutta method to integrate equations~(\ref{eq:Newton2}),
(\ref{eq:am}) and (\ref{eq:eevo}) to obtain $\bV_1$ and $\bW_1$, the
velocity and rotation rate after the first thrust.  If the star
receives a new thrust, we just select a new body frame according to
the new position of the thrust, considering $\bV_1$ and $\bW_1$ as the
initial velocity and angular velocity, and repeat the calculation as
in the first thrust and so on.

\section{Analytic Consideration}

Consider a proto-NS without initial spin. The kick consists of $n$
thrusts, each with the same duration $\tau$. We assume $n\gg1$ in this
section. After the first thrust, the star receives an angular
velocity
\begin{equation}
\Delta\Omega=\Omega_1=\frac{F\tau r \sin \alpha}{I}
 =\frac{F\tau r \sin \alpha}{kMR_{NS}^2}, \label{eq:Omega}
\end{equation}
and a velocity $\Delta V$ not more than $F\tau/M$. Note that after the
first thrust, the star's spin ${\bf\Omega_1}$ is always perpendicular
to its velocity ${\bf V}_1$. If the duration of each thrust is larger
than the spin period caused by the first thrust, i.e.
\begin{equation}
\tau \gtrsim \frac{2\pi}{\Delta\Omega}, \label{eq:aligncondition}
\end{equation}
then the second thrust will be rotationally averaged such that the net
thrust will be along $\hat {\bW_1}$ (the unit vector along
$\bW_1$). similar argument applies to additional thrusts. The final
characteristic velocity and angular velocity are given by
\begin{equation}
\ls V_f^2 \rs=\ls\lp \sum_i \Delta \bV_i \cdot {\hat \bW_1}\rp^2\rs
  \sim \frac{1}{3} n \Delta V ^2,
\end{equation}
\begin{equation}
\ls \Omega_f^2 \rs=\ls\lp \sum_i \Delta \bW_i \cdot {\hat \bW_1}\rp^2\rs
  \sim \frac{1}{3} n \Delta \Omega ^2.
\end{equation}
The typical final velocity and angular velocity are aligned, with
\begin{equation}
\bV_f \sim \sqrt{\frac{n}{3}}\Delta V {\hat \bW_1}, \quad
\bW_f \sim \sqrt{\frac{n}{3}}\Delta \Omega {\hat \bW_1}.
\end{equation}

Equation~(\ref{eq:aligncondition}) is a sufficient condition for
spin-kick alignment but not a necessary one. It's convenient to define
a critical ratio
\begin{equation}
n_c\equiv \lp \frac{2\pi}{\Delta\Omega\tau}\rp^2. \label{eq:criticalratio}
\end{equation}
For $\tau \lesssim 2\pi/\Delta\Omega$ or $n_c\gtrsim1$, spin-kick
alignment may or may not be achieved. If $n_c\gtrsim n$ or $\tau
\lesssim 2\pi/(\sqrt{n}\Delta\Omega)$, different thrusts add up in 
a random walk fashion. The final spin and kick are of order
\begin{equation}
V_f \sim \sqrt{n}\Delta V, \quad
\Omega_f \sim \sqrt{n}\Delta \Omega,
\end{equation}
with random angle between the spin and kick.

For $1\ll n_c \lesssim n$, or $2\pi/(\sqrt{n}\Delta\Omega) \lesssim
\tau \ll 2\pi/\Delta\Omega$, the situation is more complicated.
For the first $n_c$ thrusts ($i=1$, 2, ..., $n_c$), the thrust
duration $\tau$ satisfies $\tau \lesssim 2\pi/(\sqrt{i}\Delta\Omega)$,
thus the characteristic velocity and rotation rate are
\begin{equation}
V_i \sim \sqrt{i}\Delta V, \quad
\Omega_i \sim \sqrt{i}\Delta \Omega, \qquad i=1, 2, \cdots, n_c
\end{equation}
with random directions between $\bV_i$ and $\bW_i$.  For the remaining
thrusts ($i=n_c+1$, $n_c+2$, ..., $n$), $\tau \gtrsim 2\pi/\Omega_i$,
so that rotationally averaging is effective.
The final velocity and angular velocity are
\begin{eqnarray}
\bV_f &\sim& \bV_{n_c}+\sqrt{\frac{n-n_c}{3}}\Delta V {\hat \bW_{n_c}},\\
\bW_f &\sim& \bW_{n_c}+\sqrt{\frac{n-n_c}{3}}\Delta \Omega {\hat \bW_{n_c}}  
        \sim \lp \sqrt{n_c}+\sqrt{\frac{n-n_c}{3}}\rp \Delta\Omega {\hat \bW_{n_c}}.
\end{eqnarray}
Here $\bV_{n_c}\sim\sqrt{n_c}\Delta V$ has random direction compared
to $ {\hat \bW_{n_c}}$. So spin and kick will be aligned when
$\sqrt{(n-n_c)/3} \gg \sqrt{n_c}$, which means
\begin{equation}
n\gg 4n_c.
\end{equation}
Otherwise, spin and kick will be misaligned.

To summarize, spin-kick alignment/misalignment depends on the critical
ratio $n_c$ (see Eq.~\ref{eq:criticalratio}). Let $F={\cal P}/T_{\rm
kick}=M{\cal V}/T_{\rm kick}$, $r=f_\Omega R_{NS}$, we find
\begin{equation}
n_c = \lp \frac{2\pi k R_{\rm NS} n}{f_{\Omega}\tau {\cal V}\sin\alpha}\rp^2
    \simeq \lp \frac{n^2}{40 f_{\Omega} {\cal V}_{1k}T_1\sin\alpha}\rp^2,
\end{equation}
where we have used $T_{\rm kick}=n\tau$, $T_1=T_{\rm kick}/(1~{\rm s})$,
and ${\cal V}_{1k}={\cal V}/(10^3~{\rm km~s}^{-1})$. 
For 
\begin{equation}
n_c\lesssim {\rm max}(n/4, 1), \label{eq:aligncond}
\end{equation}
spin and kick will be aligned, 
while for $n_c\gtrsim {\rm max}(n/4, 1)$, spin and kick will be misaligned.

If the NS has initial spin $\Omega_{\rm init}$, the sufficient
alignment condition, equation~(\ref{eq:aligncondition}), should be modified
to
\begin{equation}
\tau \gtrsim \frac{2\pi}{{\rm max}(\Delta\Omega,\,\,\Omega_{\rm init})}.
\label{eq:alignwithinit}
\end{equation}

\section{Simulation Results}

In our model, the key parameters are $n$ and $\tau$ or $T_{\rm kick}$,
as well as the initial spin period $P_{\rm init}$.  Depending on the
kick mechanisms, the total kick duration $T_{\rm kick}$ ranges from
$0.1\,s$ to a few seconds (e.g., Lai et al. 2001; Socrates et
al. 2005; Scheck et al. 2006; Burrows et al. 2006a, b). Note that we
choose ${\cal P}=FT_{\rm kick}$ in the range of $M(500-2000){\rm
km\,s^{-1}}$ and other parameters such that the final distributions of
kick velocity and spin period of NSs qualitatively agree with
observations (see Hobbs~et al.~2005).

\begin{figure}
\centering
\includegraphics[height=15cm, angle=-90]{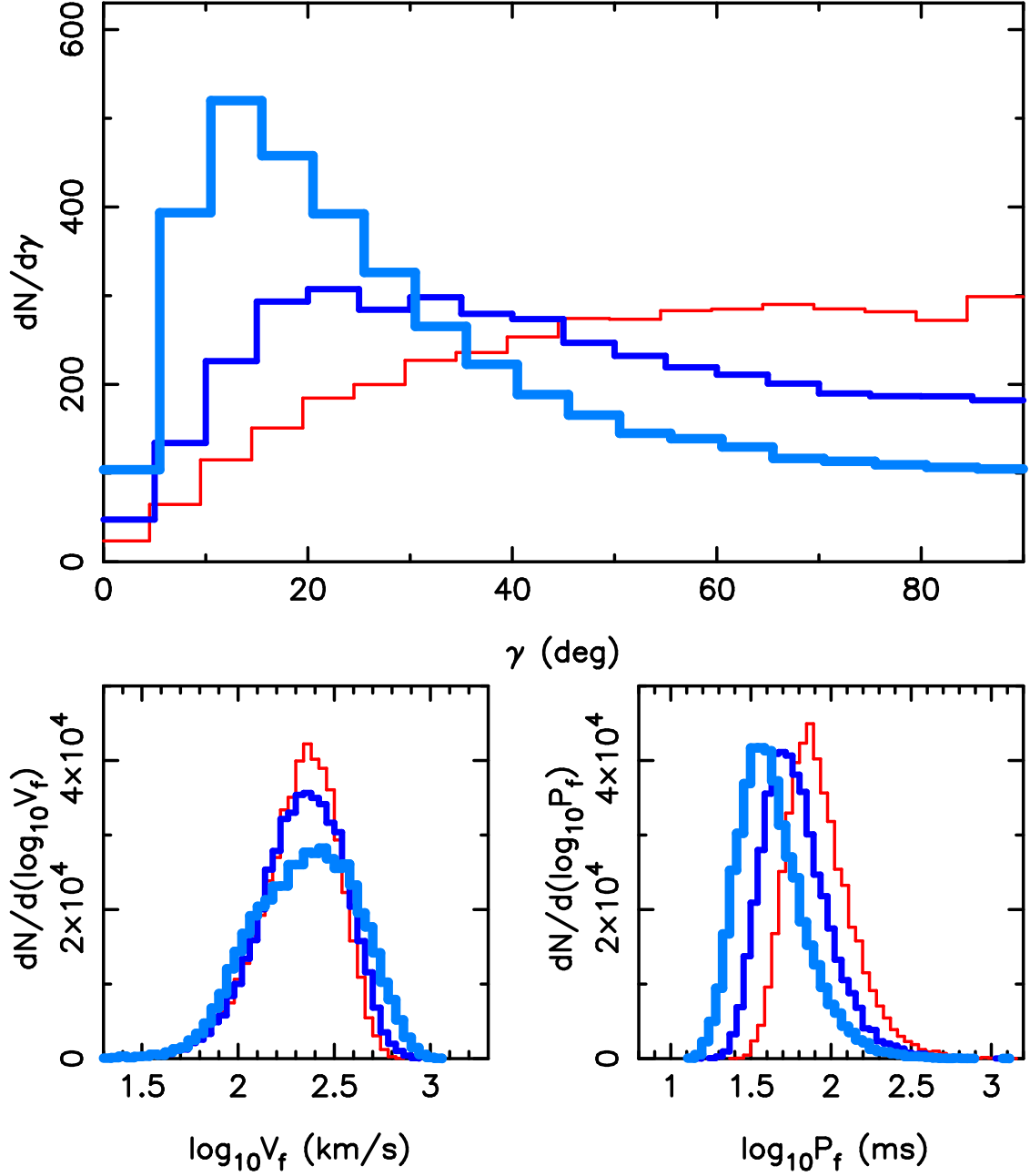}
\caption{ The distribution of the spin-kick misaligned angle $\gamma$
(upper panel), final velocity $V_f$ (lower-left panel) and final spin
period $P_f$ (lower-right panel) of NSs in simulations with $T_{\rm
kick}=1\,{\rm s}$ and different number of thrusts $n=$5, 10, 20 (from
thick lines to thin lines). The initial spin of the NS is $\Omega_{\rm
init}=0$.
\label{fig:N}
}
\end{figure}

\begin{figure}
\centering
\includegraphics[height=15cm, angle=-90]{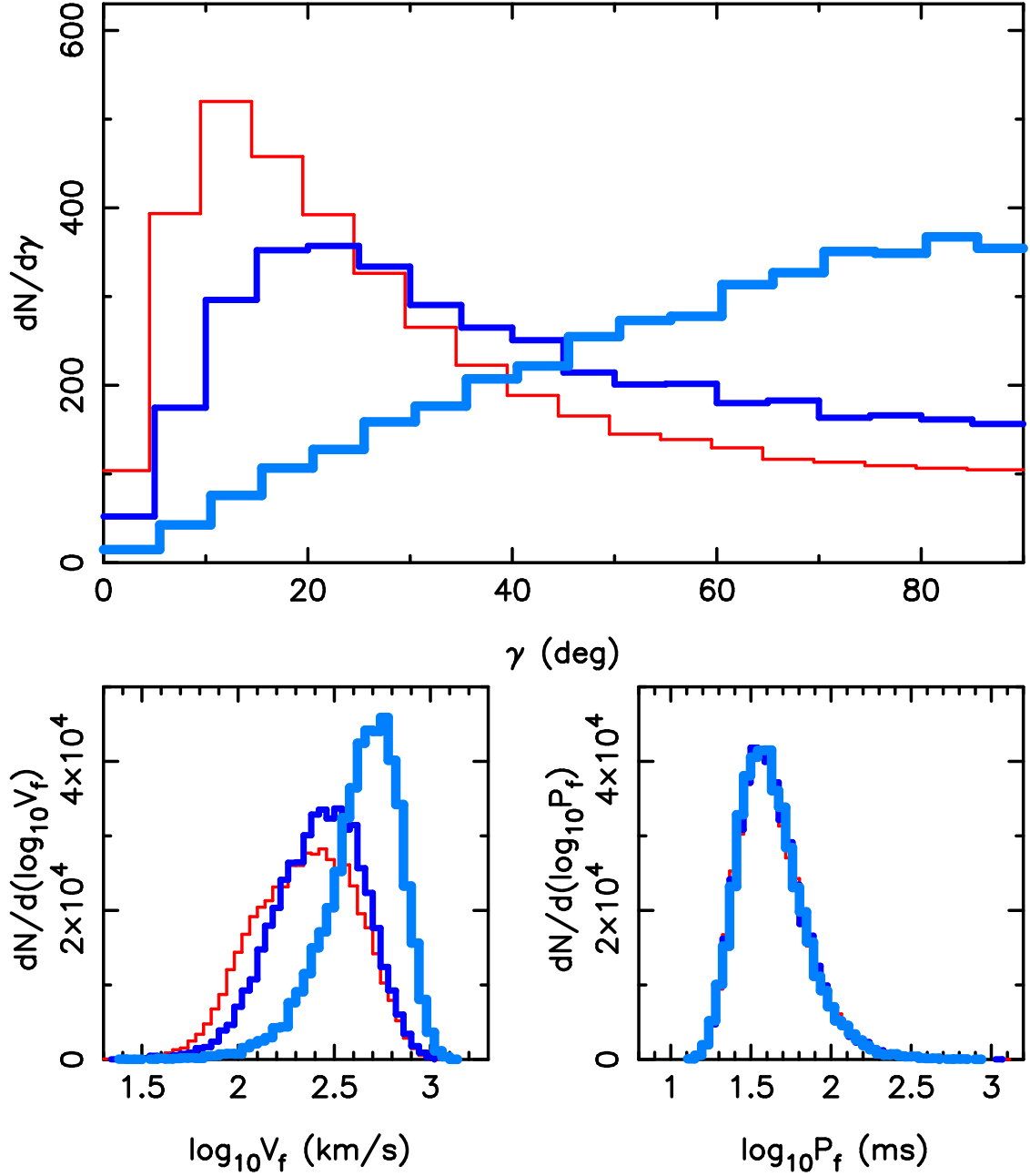}
\caption{ Same as Fig.~\ref{fig:N}, except for different values of the
total kick duration $T_{\rm kick}=$0.1s, 0.5s, 1s (from thick lines to
thin lines).  The number of thrusts is fixed at $n=5$, and the initial
spin is $\Omega_{\rm init}=0$.
\label{fig:T}
}
\end{figure}

\begin{figure}
\centering
\includegraphics[height=15cm, angle=-90]{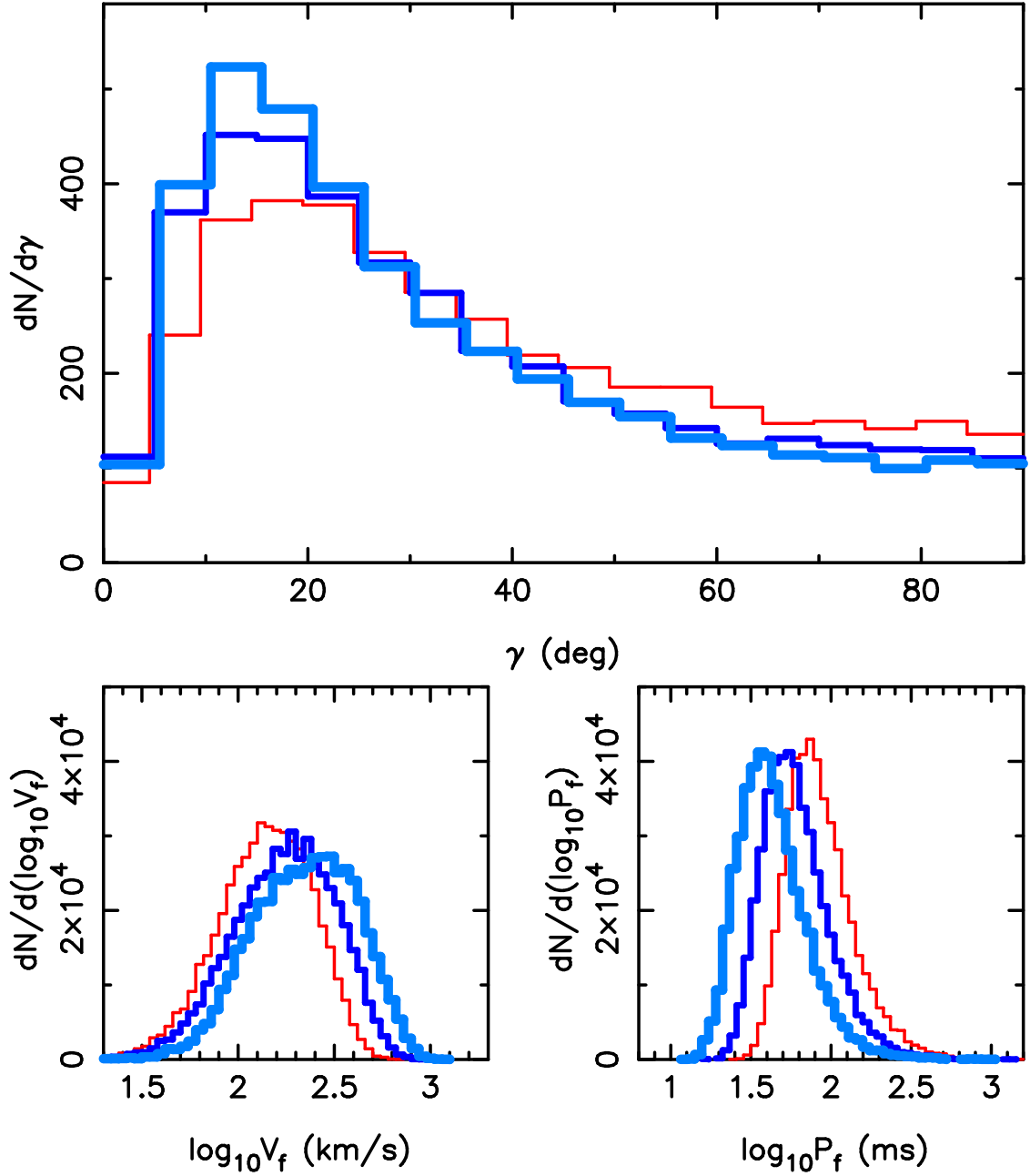}
\caption{ Same as Fig.~\ref{fig:N}, except that the 
duration of each thrust is fixed at $\tau=0.2$\,s, and 
the number of thursts are $n=5$, 10, 20
(from thick lines to thin lines). The other parameters keep
the same as Fig.~\ref{fig:N}.
\label{fig:tau}
}
\end{figure}

In Fig.~\ref{fig:N}~--~\ref{fig:tau}, we present simulations of 20000
pulsars without initial spin ($\Omega_{\rm init}=0$).  The final spins
of the NSs are all due to the off-centered thrusts.  In
Fig.~\ref{fig:N}, we fix the total kick duration to $T_{\rm
kick}=1$\,s, while changing the number of thrusts: $n=5$,~10,~20.  For
these cases, $n_c\sim \lb n/(7\sqrt{V_{1k}})\rb^4$ (see
Eq.~\ref{eq:aligncond}). We find that an aligned distribution is
produced for $n=5$, but not for $n=10$ or 20.

In Fig~\ref{fig:T}, we consider different values of total kick
duration $T_{\rm kick}=0.1$~s, 0.5~s, 1~s, while fixing the number of
thrusts to $n=5$. For these cases, $n_c\sim 0.2/(V_{1k}T_1)^2$. So we
find that for large $T_{\rm kick}$, an aligned $\gamma$ distribution
is produced.

Figure~\ref{fig:tau} shows the cases with the same thrust duration
$\tau=0.2$~s, while the number of thrusts are $n=5,~10,~20$. Here
$n_c\sim \lb n/(60V_{1k}\tau_1)\rb^2$ (where $\tau_1 = \tau /1~{\rm
s}$). Since either $n_c\lesssim 1$ or $n_c\lesssim n/4$ is
satisfied for these cases, the $\gamma$ distributions all show an
tendency of alignment.

\begin{figure}
\centering
\includegraphics[height=15cm, angle=-90]{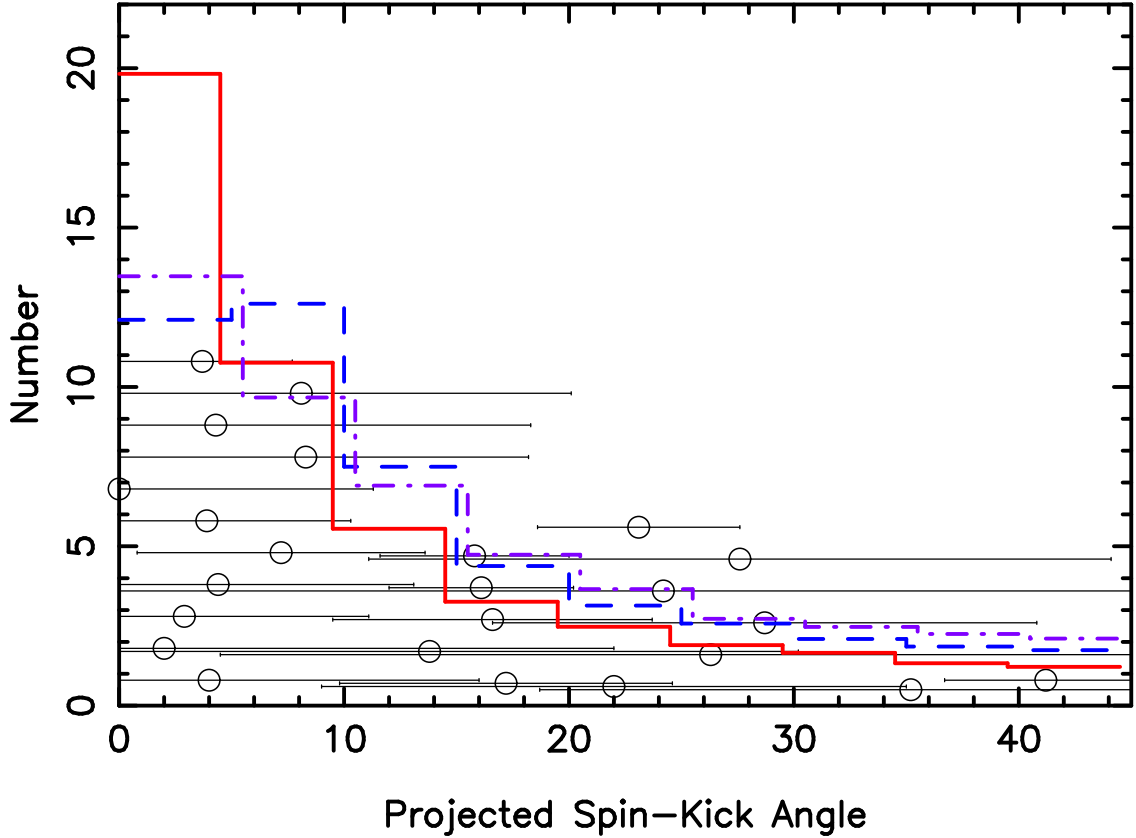}
\caption{ The spin-kick alignment distributions from observations and
simulaions. The points (with error bars) are based on radio pulsar
polarization and proper motion data (see Fig.~1 in Wang et al.~2006,
note that the spin-kick angle for the Crab pulsar has been changed
from $8\arcdeg\pm20\arcdeg$ to $26\arcdeg\pm3\arcdeg$, based on a
recent analysis, see Ng \& Romani~2006): each point represents a
pulsar with measured angle between the projected spin axis and the
proper motion direction. Note that because of the orthorgonal mode
phenomena, there is a $90^\circ$ degeneracy for the spin axis inferred
from the pulsar polarization profile.  Thus we have folded Fig.~1 of
Wang et al. around $45^\circ$ (e.g., a $60^\circ$ data point is
identified with $30^\circ$).  The histograms are based on our
simulations with $P_{\rm init}=50$~ms, $T_{\rm kick}=1$~s and $n=10$
(see Fig.~\ref{fig:P1}): The dashed line gives the actual
(unprojected) spin-kick angle distribution, the solid and dot-dashed
lines give the projected spin-kick angles in the plane of the sky when
the line of sight is inclined with respect to the initial spin axis by
$60\arcdeg$ and $30\arcdeg$, respectively. To compare with
observational data, the simuation results are also folded around
$45^\circ$.
\label{fig:obs}
}
\end{figure}

Note that in the above three figures, the kick velocities are all
distributed at a few hundred kilometers per second and the final spin
periods are distributed from 10 to hundreds of milliseconds, in
agreement with observations. Although an aligned $\gamma$
distribution can be produced under certain conditions (see
Eq.~\ref{eq:aligncond}) without initial spin, the distributions are
significantly broader than what was observed (Johnston et al.~2005;
Wang et al.~2006; see Fig.~\ref{fig:obs}).

\begin{figure}
\centering
\includegraphics[height=15cm, angle=-90]{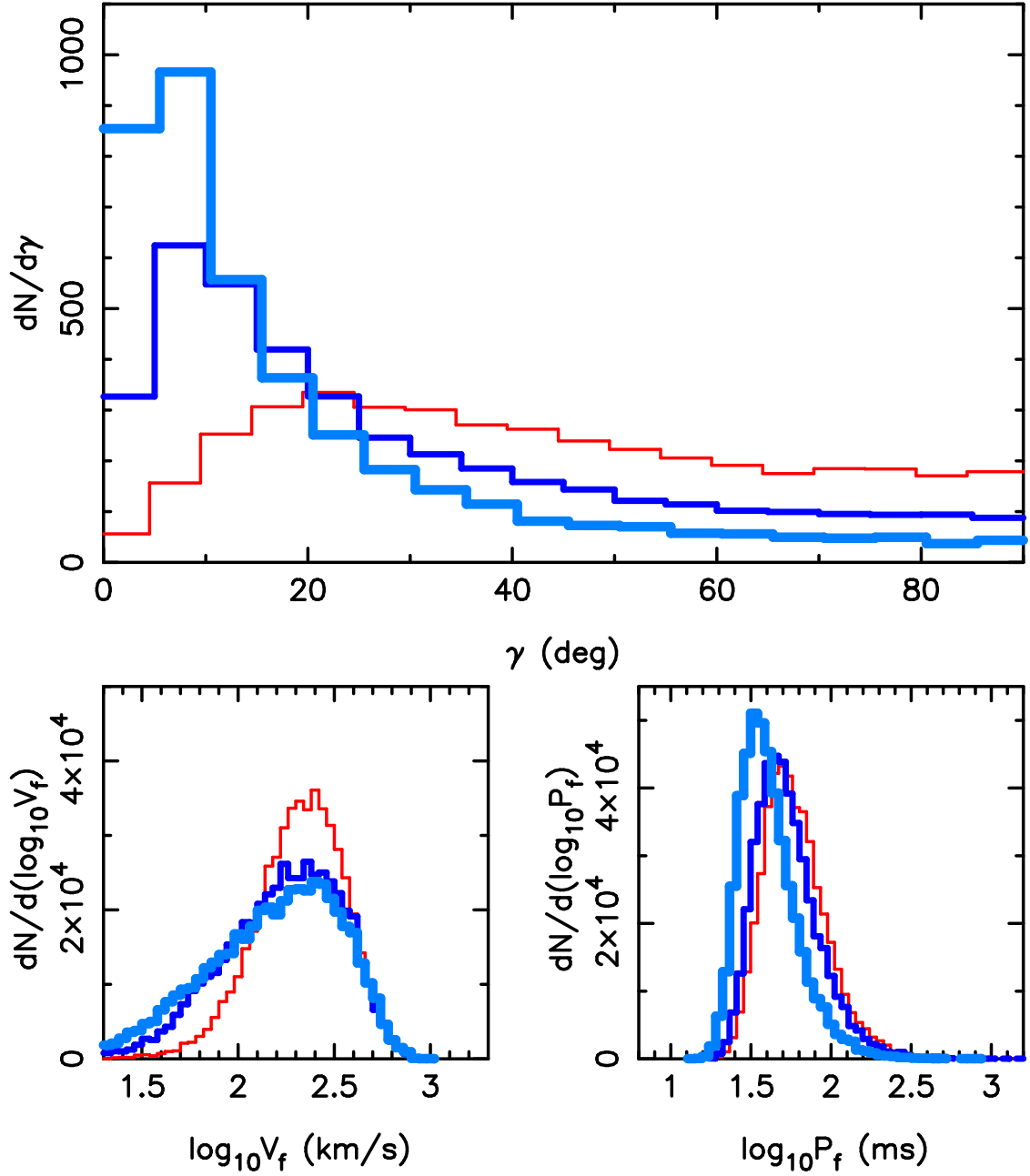}
\caption{ Same as Fig.~\ref{fig:N}, except for different values of the
initial spin period $P_{\rm init}=$50\,ms, 100\,ms, 500\,ms (from
thick lines to thin lines).  The other parameters are $T_{\rm
kick}=1\,{\rm s}$, $n=10$.
\label{fig:P1}
}
\end{figure}

Figure~\ref{fig:P1} shows the simulations with different initial spin
period $P_{\rm init}=500$\,ms, 100\,ms and 50\,ms, all with $T_{\rm
kick}=1\,$s and $n=10$. Clearly, for $P_{\rm init}\lesssim\tau$, 
rotational averaging is effective, and spin-kick alignment is easily
achieved. Figure~\ref{fig:obs} compares our simulation results
with the observed spin-kick misalignment angles based on pulsar
polarization profiles (see Wang et al.~2006), taking into account of
the sky projection effect. We see that for the $P_{\rm
init}=50$\,ms simulation depicted in Fig.~\ref{fig:P1}, the simulated 
spin-kick distribution agrees with observational data. With 
$P_{\rm init}=100$\,ms (other parameters being the same), 
the simulated distribution is broader than the data. The key condition
for producing alignment is Eq.~(\ref{eq:alignwithinit}).

\begin{figure}
\centering
\includegraphics[height=15cm, angle=-90]{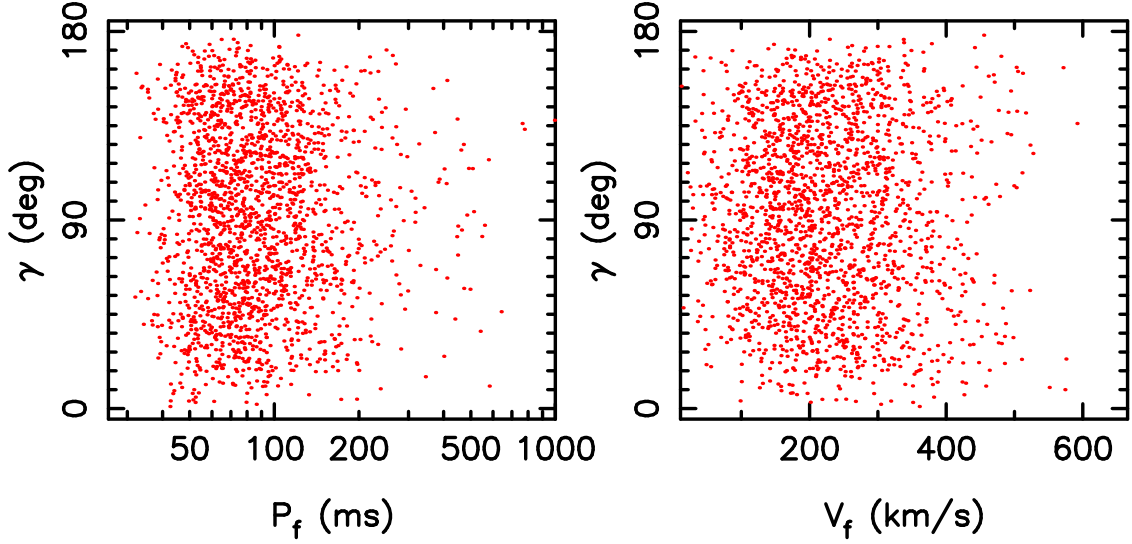}
\caption{ The distribution of the spin-kick misalignment angle
$\gamma$ as a function of the final velocity $V_f$ (left panel) and
the period $P_f$ (right panel) for a simulation which produces a broad
$\gamma$ distribution. The simulation parameters are $T_{\rm
kick}=1\,{\rm s}$, $N=20$, $\Omega_{\rm init}=0$.
\label{fig:vpg1}
}
\end{figure}

\begin{figure}
\centering
\includegraphics[height=15cm, angle=-90]{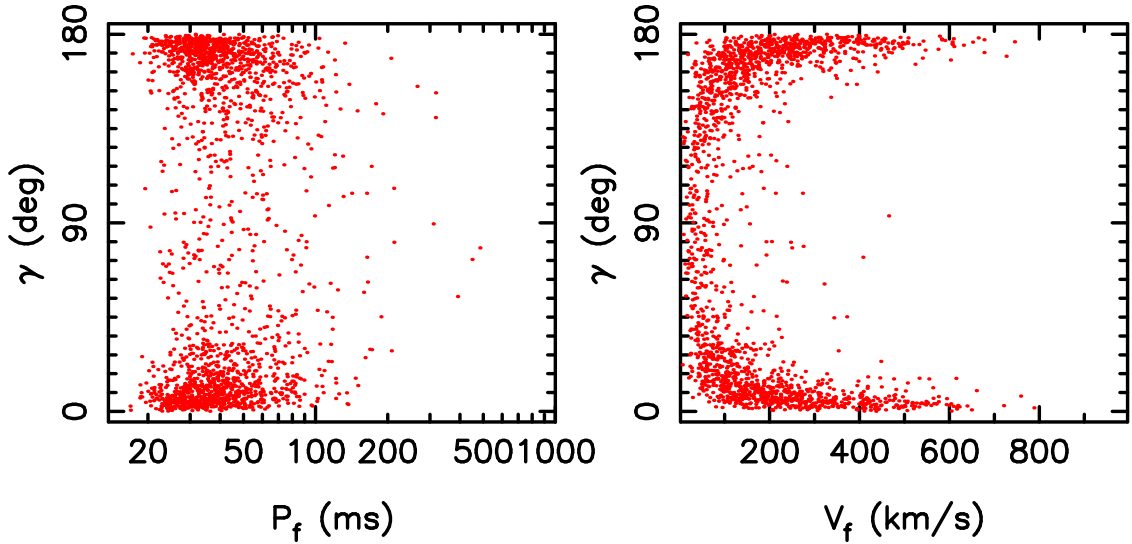}
\caption{ The distribution of the spin-kick misalignment angle
$\gamma$ as a function of final velocity $V_f$ (left panel) and period
$P_f$ (right panel) for a simulation which produces an aligned $\gamma$
distribution. The simulation parameters are $T_{\rm kick}=1\,{\rm
s}$, $N=10$, $P_{\rm init}=50$\,ms.
\label{fig:vpg2}
}
\end{figure}

Figures~\ref{fig:vpg1} and \ref{fig:vpg2} show the distribution of the
misalignment angle $\gamma$ as a function of $V_f$ and $P_f$. We see
that when the $\gamma$ distribution is broad (Fig.~\ref{fig:vpg1}),
pulsars with different $V_f$'s have similar range of $\gamma$'s.  On
the other hand, based on our toy model simulations, for an aligned
$\gamma$ distribution (Fig.\ref{fig:vpg2}), high-$V_f$ pulsars have a
strong tendency for spin-kick alignment. We have attempted to test such
$V_f - \gamma$ correlation in the existing sample for 24 pulsars. The
current data does not show such correlation, probably because of the small 
sample or large error in various measurements.

\section{Conclusion}

In this paper, we have developed a toy model to study the conditions
for pulsar spin-velocity alignment in supernova kicks.  We have
focused on the idea (Spruit \& Phinney 1998) that multiple
off-centered thrusts to the proto-neutron star may result in spin-kick
alignment.  We found that without initial angular momentum, the
alignment cannot be easily produced. To obtain the observed spin-kick
alignment distribution based on radio pulsar polarization data
(Johnston et al.~2005; Wang et al.~2006), the proton-neutron stars
should have appreciable rotation rate, with period less than the
timescale of each kick thrust. The typical initial period required is
$\lesssim 100$~ms.

Currently, the most conservative (and promising) kick mechanisms are
``hydrodynamically driven kicks''. In particular, large-scale
convections, instabilities or wave modes developed in the
neutrino-heated mantle behind the shock and in the proto-neutron star
may naturally lead to asymmetric explosion (e.g., Thompson 2000;
Scheck et al.~2004,2006; Blondin \& Mezzacappa~2006; Foglizzo et
al.~2005; Burrows et al.~2006a, b; Yamasaki \& Yamada~2006). The supernova
simulations cited above do not include initial angular momentum, and
the resulting kicks are randomly distributed. It is possible that with
even a small rotation, the hydrodynamical instabilities may
preferentially develop along the rotation axis. If the kick timescale
is long (as indicated by recent simulations), rotational averaging may
be effective and a preferentially aligned spin-kick distribution can
be produced.

Other kick mechanisms (such as those based on asymmetric neutrino
emissions in the proto-neutron star; e.g. Duncan \& Thompson 1992; Lai
\& Qian 1998; Arras \& Lai 1999a,b; Socrates et al.~2005) and the
``electromagnetic rocket'' effect (Harrison \& Tademaru 1975; Lai et
al.~2001) can easily result in spin-kick alignment, but they require
more extreme conditions (such as superstrong magnetic field or very
rapid spin) for the proto-neutron stars (see Lai 2004; Wang et
al.~2006 and references therein).

\acknowledgments

This work is supported by National Natural Science Foundation of China
(10328305, 10473015 and 10521001). D.L. has also been supported in part
by NSF grant AST 0307252 and NASA grant NAG 5-12034.  D.L. thanks NAOC
(Beijing) for hospitality during the course of the work.

\end{document}